# Spatio-temporal Attention-based Hidden Physics-informed Neural Network for Remaining Useful Life Prediction

Feilong Jiang[a]   Xiaonan Hou[a]   Min Xia[b]*
[a] Department of Engineering, Lancaster University, LA1 4YW Lancaster, U.K.
[b] Department of Mechanical and Materials Engineering, University of Western Ontario, London, Ontario, Canada
mxia47@uwo.ca

**Abstract**

Predicting the Remaining Useful Life (RUL) is essential in Prognostic Health Management (PHM) for industrial systems. Although deep learning approaches have achieved considerable success in predicting RUL, challenges such as low prediction accuracy and interpretability pose significant challenges, hindering their practical implementation. In this work, we introduce a Spatio-temporal Attention-based Hidden Physics-informed Neural Network (STA-HPINN) for RUL prediction, which can utilize the associated physics of the system degradation. The spatio-temporal attention mechanism can extract important features from the input data. With the self-attention mechanism on both the sensor dimension and time step dimension, the proposed model can effectively extract degradation information. The hidden physics-informed neural network is utilized to capture the physics mechanisms that govern the evolution of RUL. With the constraint of physics, the model can achieve higher accuracy and reasonable predictions. The approach is validated on a benchmark dataset, demonstrating exceptional performance when compared to cutting-edge methods, especially in the case of complex conditions.

## 1. Introduction

Prognostic and health management (PHM) is a technology that aims to assess the healthy state of industrial systems so to achieve optimal management [1-7]. As a critical task of PHM, Remaining Useful Life (RUL) prediction is utilized to evaluate the time remaining before repair or replacement is needed [8]. By utilizing the sensor data and operating conditions, a reasonable RUL prediction scheme can assess the condition of the industrial system to prevent accidents and reduce the cost of maintenance.

Given the rapid advancement of artificial intelligence, researchers increasingly choose deep learning (DL) methods to deal with RUL prediction tasks. Since most sensor data appears in the format of sequences, the recurrent neural network (RNN) is widely applied in the RUL prediction. Based on the Long Short-Term Memory (LSTM) network, Zhang et al. [9] proposed an architecture to estimate the working condition of the system and predict the RUL accurately. In [10], Chen et al. utilized kernel principle component analysis to obtain useful information of the input data, and the gated recurrent unit (GRU) was used for predicting RUL of complex systems. A Bidirectional Long Short-Term Memory (BiLSTM) network was proposed by Wang et al. to do the RUL estimation task [11]. The model could evaluate the sequence data forward and backward thus a higher accuracy could be obtained. Although RNN based model could evaluate the time information of sequence data effectively, the spatial information among different sensors is heavily neglected.

Thanks to its robust ability to extract spatial features, Convolutional Neural Network (CNN) has found extensive application in computer vision tasks. The combination of time sequences from different sensors can also be regarded as a 2D matrix [12]. Thus, leveraging CNN to capture the degradation feature from the 2D format data is also a popular method in the RUL prediction field. Li et al. proposed a multi-scale deep convolutional neural network structure to analyze the degradation information for reliable remaining useful life assessment [13]. Combining information from different scales, the model could make predictions in high accuracy. Jin et al. introduced the deformable convolution method into the machine RUL prediction [14]. The deformable convolution kernels could extract the information effectively and help to improve the RUL prediction capability. Meng et al. introduced a model for predicting the RUL of bearings, which combines CNN and LSTM to predict effectively [15]. However, for these methods, the information of the time scale and the importance of different sensors are ignored.

Owing to its ability of paying more attention to important information [16], the attention mechanism has made fundamental influence in the deep learning field. Researchers have tried to incorporate the attention mechanism into the RUL prediction task to assign larger weight to more useful data. A feature-attention mechanism was proposed by Liu et al., which was used to find more useful sensor data [17]. This mechanism could help the model focus more on more important input and improve accuracy. Zhang et al. adopted the Transformer structure for RUL prediction. Thanks to the self-attention mechanism, this method could evaluate the input data effectively and make accurate predictions. Li et al. introduced a multi-task spatio-temporal network for predicting RUL [18]. The attention strategy on both the local and global scale helped to enhance the prediction capability of RUL.

Although data-driven machine learning approaches have achieved certain success in the RUL prediction field, the lack of interpretability remains unsolved, which prevents the implementation of machine learning in PHM. Hence, there has been a notable surge in interest in physics-based RUL prediction in recent years. By analyzing the nature of the system degradation, the traditional physics-based RUL prediction approach directly builds mathematical models based on the priori engineering knowledge of researchers. However, owing to the complexity of industrial systems, obtaining high fidelity model with good generalization ability is almost impossible [19] which heavily prevents its implementation. With the fast development of physics informed machine learning [20], it leads to new approaches to realize the physics-based RUL prediction. By integrating the simplified physical equations [21] or hidden physics laws [22] into the machine learning model, the models will be trained by the combination of data and physics constraints. Under such circumstances, the physics informed machine learning model can use the training data more effectively and at the same time, make more reasonable predictions following the underlying physics. Considering the irreversible character of RUL, Chen et al. introduced a physical loss term into the loss function to ensure the monotonicity of the degradation process [23]. However, the model overlooked the physical connection between the sensor data and the RUL, thus constraining the interpretability of the results. Utilizing the Deep Hidden Physics Model (DeepHPM) to capture the mechanisms that govern the evolution of RUL, Cofre-Martel et al. successfully improved the accuracy of their basic model [24]. However, the model failed to account for the varying significance of input data, thus limiting its ability to effectively extract features. Liao et al. applied the self-attention mechanism to the feature extraction network and DeepHPM to improve the RUL prediction capability [25]. Although the method showed an improvement in the prediction accuracy, time information in the input data was almost neglected.

To resolve the crucial issues mentioned above in RUL prediction, we propose a Spatio-temporal Attention-based Hidden Physics-informed Neural Network (STA-HPINN). In this method, the self-attention mechanism is employed to allocate weights to various sensors as well as time steps to capture the spatio-temporal information. Adapted from DeepHPM [22], the self-attention-based Hidden Physics Informed Neural Network (AHPINN) is employed to distill the hidden physics relationship between the input data and the corresponding RUL.

The primary contributions of this paper are outlined as follows.

1) A novel spatio-temporal attention model is proposed for predicting RUL. The self-attention mechanism [26] is employed to find more useful sensors and time steps. By combining the outcome of the sensor attention and time step attention, our model can extract the degradation information more effectively.

2) AHPINN is used to make predictions and capture the hidden physics between the sensor data and RUL. Benefiting from the physical constraint, the model can make reasonable predictions with improved interpretability and accuracy than pure data-driven methods.

The structure of the rest of this paper is as follows. Section 2 discusses the methodology. Then we present the experiment on the C-MAPSS dataset in section 3 [27]. Section 4 talks about the ablation study. The conclusion is drawn in section 5.

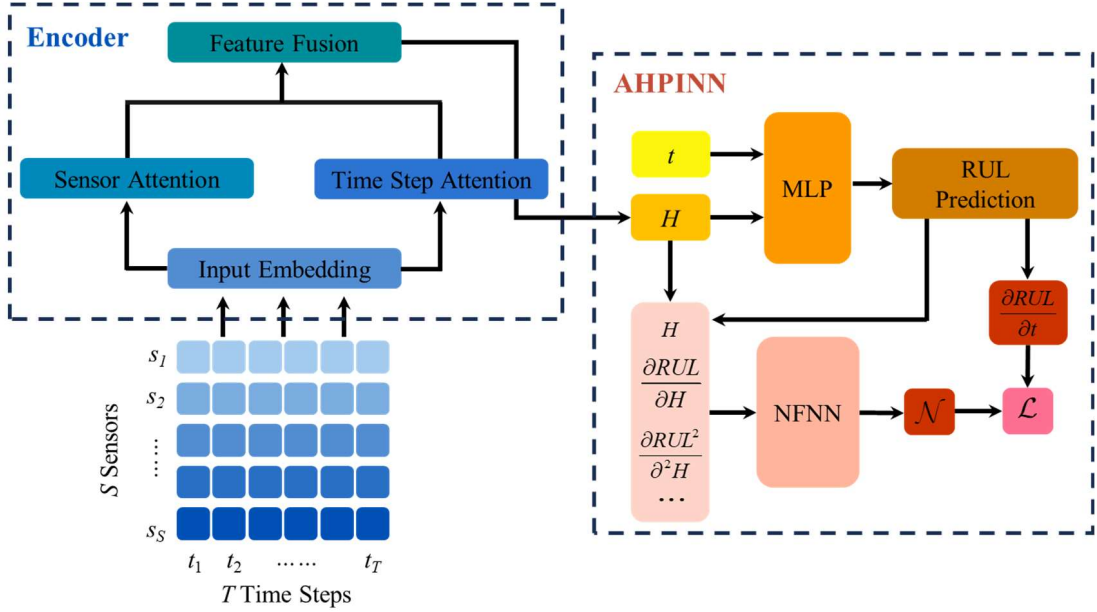

**Fig. 1.** Proposed Spatio-temporal Attention-based Hidden Hhysics-informed Neural Network.

## 2. Methodology

In this section, we present a detailed explanation of both the RUL prediction problem and the proposed framework.

2.1. Problem Description

The condition monitoring (CM) data (e.g. temperature, vibration, fan speed etc.) collected by various sensors during the operation of a system are used to predict the RUL. Utilizing the input data $X \in \mathbb{R}^{T \times S}$, the RUL prediction task can be expressed as:

$$\text{RUL} = f(X, \theta), \tag{1}$$

where $X$ is time series data during the operation of the machine, $T$ denotes the number of time steps, and $S$ represents the number of sensors. $f$ represents the prediction model, $\theta$ denotes the parameters of the model (e.g. weights and biases).

The proposed model's structure, as illustrated in Fig. 1, primarily comprises two components: an encoder and the AHPINN. The details of the architecture and main blocks are presented as follows.

## 2.2. Encoder

To better utilize the input data and reduce redundancy, an encoder is employed to extract degradation information and decrease the dimension of the input data. The encoder is composed of 3 blocks, which are input embedding block, attention block and feature fusion block. A detailed explanation of these blocks is presented below.

### 2.2.1. Input Embedding

To deal with input data with different dimensions, an embedding block is used to unify the dimensions of the input and make preparation for the feature fusion process. The input embedding transforms the input data into a $D$-dimensional vector by a feed forward linear layer. For the sensor attention block, the embedded dimension is $S \times D$. For the time step attention block, the embedded dimension is $T \times D$.

### 2.2.2. Sensor and Time Steps Attention

The self-attention mechanism is employed to calculate attention scores for various sensors across different time steps. A sensor attention block is utilized to extract features from the sensor dimension. The time step attention block is responsible for extracting the feature from the time step dimension. The attention blocks for sensors and time steps share the same structure. As shown in Fig. 2, the attention block contains 2 sublayers, which are a self-attention layer and a fully connected feed-forward layer. The residual connection is employed in each sublayer with layer normalization operation.

The input of the self-attention layer contains query ($Q$), key ($K$) of dimension $d_k$, and value ($V$) of dimension $d_v$. The $Q$, $K$ and $V$ are all obtained by passing the input data through a linear layer. The self-attention mechanism can be expressed as:

$$Attention(Q,K,V) = \text{softmax}\left(\frac{QK^T}{\sqrt{d_k}}\right)V. \qquad (2)$$

$Attention(Q,K,V)$ represents the attention score matrix. $\sqrt{d_k}$ is used to scale the dot products to prevent $QK^T$ falls into the area where the softmax function has extremely small gradients.

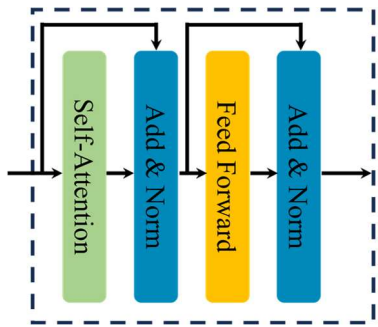

**Fig. 2** Structure of attention block.

### 2.2.3. Feature Fusion

The feature fusion block is responsible for combining the features of sensor attention and time step attention and transforming the new feature into a low-dimensional representation, which contains the compressed information from the original sensor data. The structure of the feature fusion block is shown in Fig. 3. First, the outputs of the sensor attention block $F_s \in \mathbb{R}^{S \times D}$ and the outputs of the time step attention block are concatenated to get a new feature map $F_r \in \mathbb{R}^{(T+S) \times D}$. Then, a convolutional layer with a kernel size of $(T+S) \times 1$ is employed to extract the features from $F_r$ and fuse the information from sensor attention block and time step attention block. The stride is set as 1, and the output channels are 3. The squeeze and excitation network (SENet) [28] is used to assign weight to different channels. The outputs are then flattened and fed into a linear layer to get the low-dimensional hidden state ($H$). The dimension of the hidden state is 3.

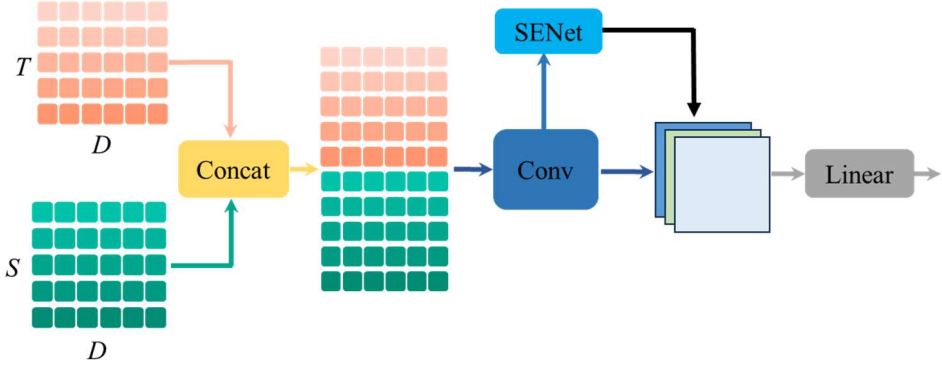

Fig. 3 The proposed feature fusion structure.

2.3. AHPINN

Considering the fact that the physical relationship (explicit partial differential equations) between sensor data and the corresponding RUL remains unclear, directly integrating the physical knowledge into the neural network is hard to implement. To overcome this problem, AHPINN is implemented to capture the hidden physical connections between the sensor data and the RUL.

The inputs of AHPINN are the hidden state of the sensor data and the corresponding time (t). The output is the corresponding RUL. The working principle of AHPINN is described as follows.

Even though we don't know the exact PDEs between sensor data and the corresponding RUL, we can still roughly describe its form by the general form of partial differential equations (PDEs) as [21], [22]:

$$\frac{\partial u}{\partial t} - \mathcal{N}(u) = 0, \tag{3}$$

$u$ represents the latent solution, which corresponds to RUL in RUL prediction task. $\mathcal{N}$ denotes the nonlinear function. $\mathcal{N}$ can be expressed in detail as (assuming the highest order of derivative is 2):

$$\begin{aligned}
\mathcal{N}(u, u_x, u_{xx}) = & \alpha_{0,0} + \alpha_{1,0} u + \alpha_{2,0} u^2 + \alpha_{3,0} u^3 + \\
& \alpha_{0,1} u_x + \alpha_{1,1} u u_x + \alpha_{2,1} u^2 u_x + \alpha_{3,1} u^3 u_x + \\
& \alpha_{0,2} u_{xx} + \alpha_{1,2} u u_{xx} + \alpha_{2,2} u^2 u_{xx} + \alpha_{3,2} u^3 u_{xx},
\end{aligned} \tag{4}$$

where $\alpha$ is hyper parameter. Although for the RUL prediction problem, $\mathcal{N}$ is unclear, one can still try to capture the hidden physics relationship between sensor data and the RUL by neural network. Here we introduce AHPINN. The AHPINN is adapted from the original work of DeepHPM [22] with the self-attention mechanism to enhance its capability.

As shown in Fig. 1. The AHPINN is composed of two networks. The first network, which is a multilayer perceptron (MLP), is responsible for predicting the RUL with $H$ and $t$ as inputs. The second network, which is a nonlinear function neural network (NFNN), is used to distil the hidden mechanisms

(nonlinear function, $\mathcal{N}$ in equation (4)) between the compressed sensor data $H$ and RUL. The inputs of NFNN are $\left[\dfrac{\partial RUL}{\partial H}, \dfrac{\partial^2 RUL}{\partial H^2}, \cdots\right]$, which are the derivatives of RUL with respect to $H$. The maximum order of the RUL derivative with respect to $H$ is set to 3 [25]. Thus, the output of NFNN is $\mathcal{N}\left(H, \dfrac{\partial RUL}{\partial H}, \dfrac{\partial^2 RUL}{\partial H^2}, \dfrac{\partial^3 RUL}{\partial H^3}\right)$. In the original work of DeepHPM, the second network is also an MLP. Here we apply the self-attention mechanism to enhance its feature extraction ability. The structure of the attention based NFNN is shown in Fig. 4.

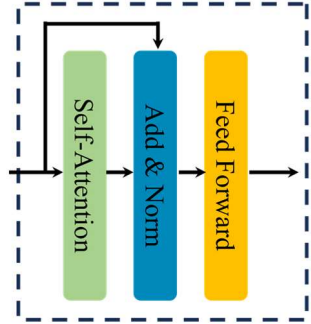

**Fig. 4.** The structure of attention based NFNN.

The loss function of the whole model is formed by two terms, which are the data driven loss term and physics driven loss term. The loss function can be expressed as:

$$\mathcal{L} = \lambda_1 \mathcal{L}_{data} + \lambda_2 \mathcal{L}_f, \qquad (5)$$

where $\lambda$ is the weight factor, which is used to balance different terms. In this paper the ReLoBRaLo scheme [29] is used to adjust the value of $\lambda$. $\mathcal{L}_{data}$ is data driven loss term, which can be expressed as:

$$\mathcal{L}_{data} = \dfrac{1}{N}\sum_{i=1}^{N}\left(RUL_i - \widehat{RUL_i}\right)^2, \qquad (6)$$

where $\widehat{RUL}$ denotes the predicted RUL, RUL and denotes the real RUL. $\mathcal{L}_f$ is the physics driven loss term:

$$\mathcal{L}_f = \dfrac{1}{N}\sum_{i=1}^{N} f(H_i, t_i), \qquad (7)$$

where $f$ is:

$$f = \dfrac{\partial RUL}{\partial t} - \mathcal{N}. \qquad (8)$$

Apparently, different from pure data driven methods, the training process not only need to minimize the data drive loss term but also the physics driven loss term. The process of minimizing the physics driven loss term $\mathcal{L}_f$ equals to solve the equation in the form of equation (3). In this way, the training of the model is a process of constructing a mapping relationship between the sensor data and RUL, while simultaneously approximating the nonlinear function linking the sensor data and RUL. Consequently, the model is trained by both the data and physics constraints. With the embedding of the physics, the model can achieve higher accuracy and make reasonable prediction.

## 3. Experiment

In this section, we validate the performance of our model on the C-MAPSS dataset. Data processing and experimental setting are described in detail. Comparisons with the cutting-edge approaches are presented.

3.1. Data Processing

The benchmark dataset in RUL prediction, the C-MAPSS dataset provided by NASA, is chosen for experiments. The C-MAPSS dataset comprises 4 subsets, each containing simulated sensor data depicting the progression from a healthy state to the failure state of aircraft engines under various operational conditions and failure modes. Every subset is split into a training set and a testing set. More information on each subset can be found in Table 1.

**Table 1**
Information of the C-MAPSS dataset

|  | FD001 | FD002 | FD003 | FD004 |
|---|---|---|---|---|
| Operation Conditions | 1 | 6 | 1 | 6 |
| Fault Modes | 1 | 1 | 2 | 2 |
| Training Trajectories | 100 | 260 | 100 | 248 |
| Testing Trajectories | 100 | 259 | 100 | 249 |

It's noticeable that FD001 is the simplest subset, featuring only 1 operation condition and 1 fault mode. In contrast, FD004 emerges as the most complex subset, encompassing 6 operation conditions and 2 fault modes.

The data of sensors T2, P2, P15, epr, farB, Nf_dmd, and PCNfR_dmd stay constant during the whole process, thus the corresponding data are removed. Min–max normalization is implemented to normalize the original data. Previous research shows that it is more appropriate to set the truncation threshold of RUL to 125 [30] which is also adopted in this paper.

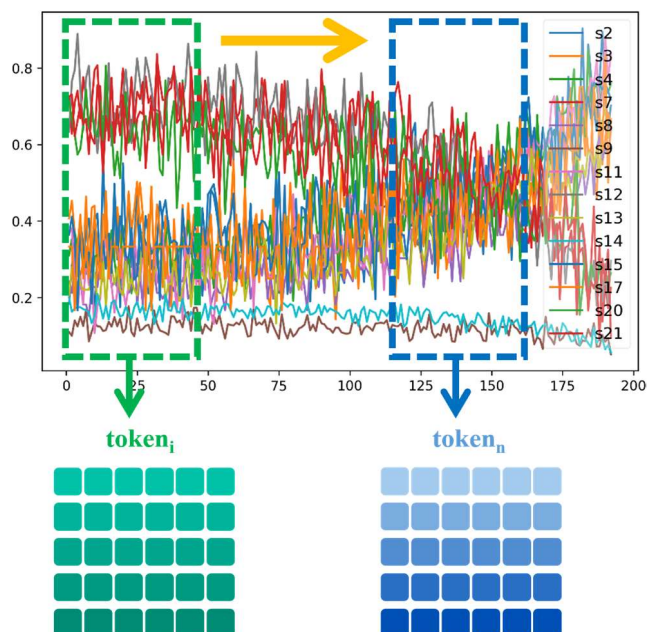

**Fig.5** The sliding window process.

The sliding window process, which is a commonly used data segmentation process to get the time information from the time series data, is used to segment the sensor data. The diagram of the sliding window process is shown in Fig. 5. The RUL of the last component of each token is set as the label. The window sizes are set as 40 for FD001 and 60 for the others. This is because FD002, FD003 and FD004 have more complex operation conditions and fault modes than those of FD001. A larger window size means more degradation information is contained in one token. It aids in improving the capability of predicting RUL in complex conditions. The sliding stride is 1.

3.2. Experimental Setting

In this work, 20% of the original training data are randomly chosen as the validation data set. The validation data set is used to monitor the training process to obtain the best result. To reduce the difference in magnitude between input and output and accelerate the training process, the output value of the model is multiplied by 125. The batch size is set as 512. In this model, the MLP in AHPINN contains 3 hidden layers with 10 neurons per layer. Before each activation function, batch normalization is used to normalize the output of the previous layer. Tanh is adopted as the activation function. Adam optimizer is used to update the model. The learning rate is set as 0.001 for the first 50 iterations, and then 0.0001. The training is conducted on a single NVIDIA GeForce RTX 4090 GPU.

We employ the RMSE and scoring function to assess the performance of our model. The RMSE quantifies the average deviation between the prediction and the actual value. It can be displayed as:

$$RMSE = \sqrt{\frac{1}{N}\sum_{i=1}^{N}\left(RUL_i - \widehat{RUL_i}\right)^2}. \qquad (9)$$

The scoring function penalizes delayed predictions more heavily. The scoring function can be expressed as:

$$s = \sum_{i=1}^{N} s_i, \; s_i = \begin{cases} e^{-\frac{d_i}{13}} - 1, \text{for } d_i < 0 \\ e^{\frac{d_i}{10}} - 1, \text{for } d_i \geq 0 \end{cases} \qquad (10)$$

where $d_i = \widehat{RUL_i} - RUL_i$.

The final result is obtained by averaging 10 independent trials.

3.3. Result and Comparison with Other Methods

**Table 2**

Comparison with existing methods on the C-MAPSS dataset

| Approach | RMSE | | | | | Score | | | | |
|---|---|---|---|---|---|---|---|---|---|---|
| | FD001 | FD002 | FD003 | FD004 | Average | FD001 | FD002 | FD003 | FD004 | Average |
| DCFA (2023) [31] | 11.74 | 16.81 | 10.71 | 17.77 | 14.26 | 190 | 1076 | 198 | 1571 | 758.75 |
| 3D Attention (2023) [32] | 13.12 | 13.93 | 12.15 | 20.34 | 14.85 | 231.01 | 759.84 | 195.56 | 1710.29 | 724.18 |
| MTSTAN (2023) [18] | **10.97** | 16.81 | 10.90 | 18.85 | 14.38 | **175.36** | 1154.36 | 188.22 | 1446.29 | 741.06 |
| DVGTformer (2024) [33] | 11.33 | 14.28 | 11.89 | 15.50 | 13.25 | 179.55 | 797.26 | 254.55 | 1107.5 | 584.72 |
| ChangePoint-LSTM (2024) [34] | 13.59 | 16.67 | 12.94 | 18.69 | 15.47 | 224.88 | 947.99 | 207.10 | 1360.34 | 685.08 |
| Optimizing Distribution Parameters (2024) [35] | 11.96 | 13.51 | 11.40 | 17.58 | 13.61 | 233.40 | 902.13 | 255.6 | 1704.59 | 773.93 |
| STA-HPINN (Proposed) | 11.27 | **13.21** | **8.30** | **13.31** | **11.52** | 211.03 | **764.42** | **119.93** | **849.98** | **486.34** |

**Bold** means the best result in all methods.

Table 2 shows the comparison with the cutting-edge approaches. Except for FD001, the performances of our method show significant improvement compared with existing approaches. Taking into account that the complexity of operational conditions and fault modes in FD002, FD003, and FD004 are more complex, it indicates that with the aid of physics constraint, our model has superior ability for RUL prediction under complex conditions.

Fig. 6 displays the test results of the proposed model across the 4 subsets. The predicted RUL closely aligns with the actual RUL for each subset. This shows the efficacy of the proposed approach.

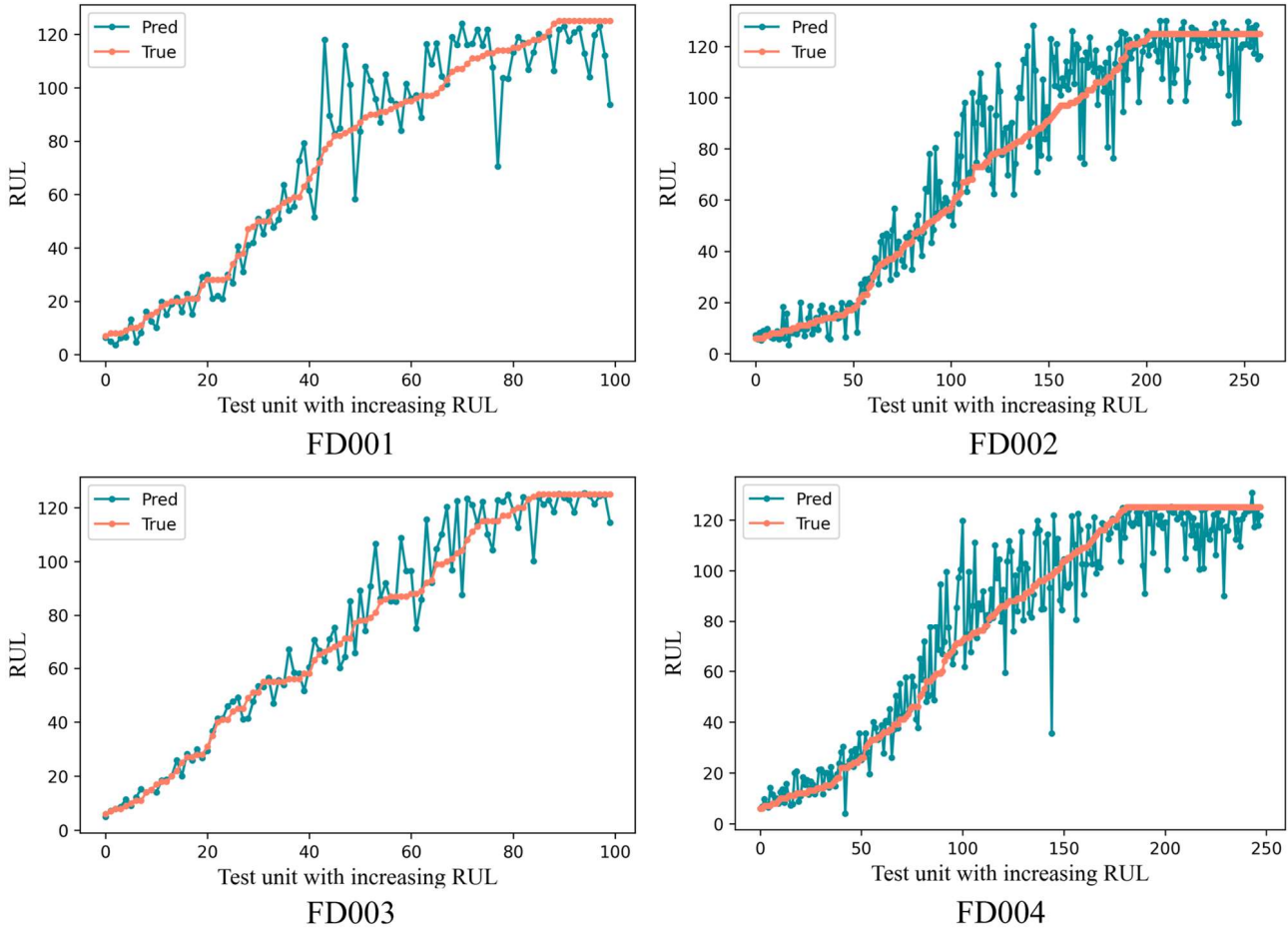

**Fig.6.** True RUL and the predicted RUL by the proposed method

## 4. Ablation Study

This section shows the outcomes of the ablation study along with the corresponding discussion.

### 4.1. Ablation of AHPINN

Table 3 presents the comparisons of the model with AHPINN and without AHPINN. It is evident that except for the RMSE on subset FD003, the model with AHPINN shows better performance. Notice that for the subsets FD002 and FD004, which have complex operation conditions, the model with AHPINN shows significant improvement. This indicates that the proposed method, which integrates physics constraints, is more effective.

To provide additional insights, we visualize the hidden vectors extracted by the encoder with the real RUL as the label. Fig. 7 shows the hidden vectors of the test data set of FD002 in 3D space and the projection on feature $x_2$ and $x_3$. It can be observed that for the model with AHPINN, the hidden vectors show a strong linear distribution. It clearly shows that with the physics constraint, the vectors sharing similar RUL cluster together and the projection on $x_2$ and $x_3$ shows a clear degradation direction which explains why the model with AHPINN can obtain higher accuracy.

**Table 3**
Comparison of the prediction performance of the model with AHPINN and without AHPINN

| Subsets | RMSE | | Score | |
| --- | --- | --- | --- | --- |
| | AHPINN | Without AHPINN | AHPINN | Without AHPINN |
| FD001 | **11.27** | 11.47 | **211.03** | 232.29 |
| FD002 | **13.21** | 13.84 | **764.42** | 992.43 |
| FD003 | 8.30 | **8.26** | **119.93** | 121.17 |
| FD004 | **13.31** | 14.26 | **849.98** | 1029.42 |
| Average | **11.52** | 11.96 | **486.34** | 593.83 |

**Bold** means the best result in all methods.

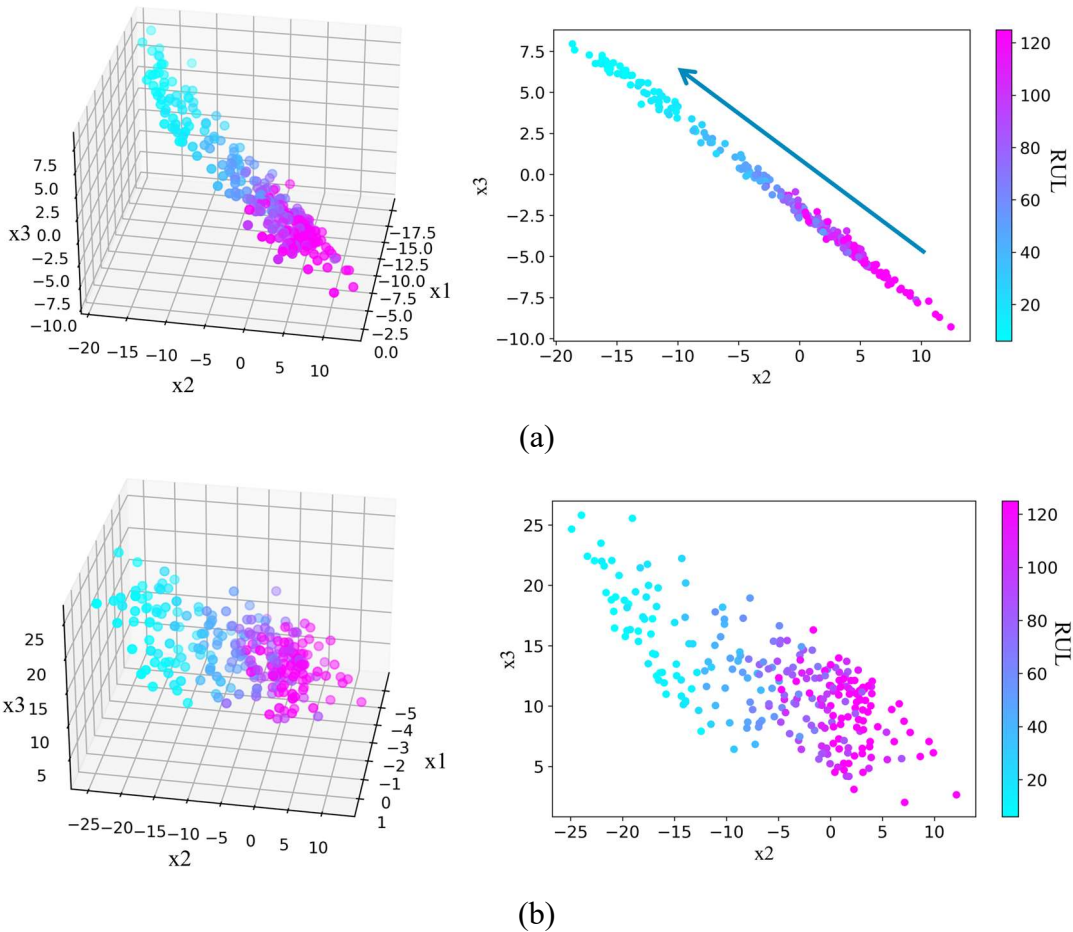

(a)

(b)

**Fig.7.** Visualized hidden vectors of FD002 test data set. (a) Model with AHPINN. (b) Model without AHPINN.

To further analyze, we visualize the hidden vectors of a specific engine of the FD004 test data set. Fig. 8 shows the hidden vectors of test engine #213. It is evident from Figure 8 (a) that the hidden vectors cluster into 3 distinct groups. As marked by the arrows in the RUL prediction curve of the model without AHPINN, there are two apparent breaks of the predicted RUL curve. It is evident in the first figure of Fig. 8 (a) that the breaks occur at the cycles when the model makes predictions from one group to another. This is because the groups are separated from each other, and the information between the adjacent groups is missing. Without enough degradation information, the model cannot make predictions reasonably. It can also be observed from the first figure of Fig. 8 (a) that for the hidden vectors around cycle 200, the predicted RULs get similar values without degrading. This can be explained by the distribution of group 2. For group 2, the hidden vectors corresponding to different real RULs are overlapped together, which makes it hard for the model to distinguish them from each other and causes inaccurate predictions. As shown in Fig. 8 (b), the model with AHPINN can largely solve the problems mentioned above.

The examples show that owing to the regulation of hidden physics relationships, the encoder can extract the hidden features more reasonably. Improved distribution of the hidden vectors leads to more precise RUL prediction.

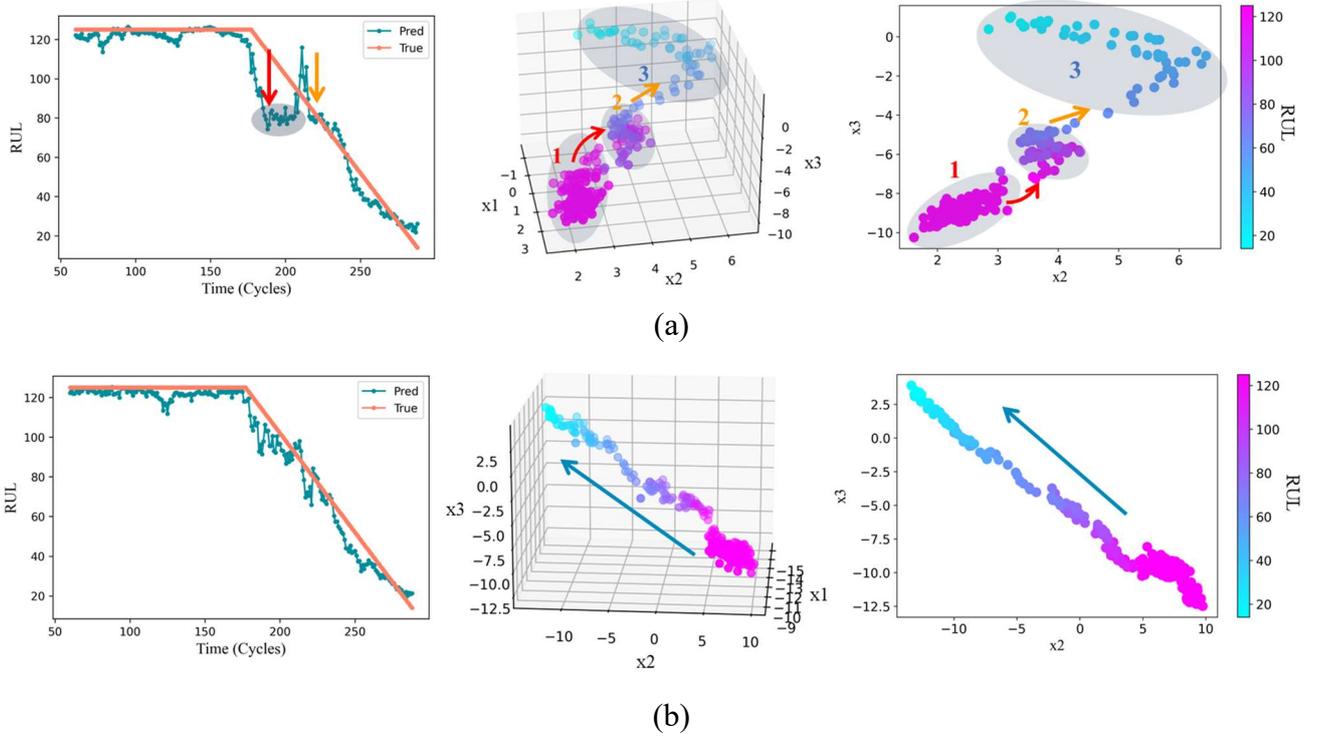

(a)

(b)

**Fig.8.** The predicted RUL curve and hidden vectors' distribution of test engine #213. (a) The predicted RUL curve and hidden vectors' distribution of the model without AHPINN. (b) The predicted RUL curve and hidden vectors' distribution of the model with AHPINN.

4.2. Ablation of Attention Mechanism

Table 4 shows the results of the ablation study of time step attention (M1), sensor attention (M2) and the attention mechanism in AHPINN (M3). Notice that for M1, the model shows better performance on FD001 and FD003. As for M2, it achieves better accuracy on FD002 and FD004 but shows worse performance on FD001 and FD003 than M1. By combining the features of time step attention and sensor attention, the proposed method successfully inherits the advantage of M1 and M2. Consequently, it achieves better overall accuracy than M1 and M2. It can be seen that M3 shows worse performance of all 4 subsets compared to STA-HPINN. It means the attention mechanism in AHPINN helps to enhance the capability of the proposed model.

**Table 4**
Ablation study of attention mechanism

| Approach | RMSE | | | | | Score | | | | |
|---|---|---|---|---|---|---|---|---|---|---|
| | FD001 | FD002 | FD003 | FD004 | Average | FD001 | FD002 | FD003 | FD004 | Average |
| M1 | **11.09** | 17.73 | 8.55 | 21.95 | 14.83 | 212.38 | 2530.58 | 133.74 | 4503.22 | 1844.98 |
| M2 | 11.71 | 13.62 | 9.73 | 13.56 | 12.16 | 240.41 | 839.74 | 171.02 | 871.20 | 530.59 |
| M3 | 11.34 | 13.39 | 8.89 | 13.70 | 11.83 | 220.42 | 822.03 | 138.59 | 900.95 | 520.50 |
| STA-HPINN | 11.27 | **13.21** | **8.30** | **13.31** | **11.52** | **211.03** | **764.42** | **119.93** | **849.98** | **486.34** |

**Bold** means the best result in all methods.

## 5. Conclusion

In this paper, a Spatio-temporal Attention-based Hidden Physics-informed Neural Network was developed for RUL prediction. The spatio-temporal attention mechanism was proposed to emphasize the sensors and time steps that contain more useful degradation information. The extracted features were fused to remove redundant features and get the hidden vectors that contain useful information. An attention based AHPINN was used to predict RUL and capture the hidden physics between the hidden state and the RUL. With such a method, our model showed outstanding performance with strong interpretability. Owing to the physics constraint, our model achieved superior capability in RUL prediction tasks than cutting-edge methods, especially under complex conditions. The proposed method has significant potential in practical applications where model reliability and interpretability are crucial.